\documentclass[iop,apj]{emulateapj}
\usepackage{graphicx}
\usepackage{amsmath,amssymb}

\begin{document}
\title{Ion collisional transport coefficients in the solar wind at 1 AU}

\author{Petr Hellinger}
\email{petr.hellinger@asu.cas.cz}

\affiliation{Astronomical Institute, CAS,
Bocni II/1401, CZ-14100 Prague, Czech Republic
}

\begin{abstract}
Proton and alpha particle collisional transport coefficients (isotropization, relative deceleration 
frequencies and heating rates)
at 1 AU are quantified using the WIND/SWE data. In agreement with
previous studies the ion-ion Coulomb collisions are generally important for
slow solar wind streams and
tend to reduce the temperature anisotropies,
the differential streaming and the differences between proton and alpha particle
temperatures. 
In slow solar wind streams the Coulomb collisions between protons and alpha
particles are important for the overall proton energetics 
as well as for
the relative deceleration between the two species.
It is also shown 
that ion temperature anisotropies and differential streaming
need to be generally taken into account for evaluation of the collisional transport coefficients.
\end{abstract}
\pacs{?}
\maketitle

\section{Introduction}

The solar wind plasma is weakly collisional and far from thermal equilibrium;
the observed particle velocity distribution functions exhibit important temperature
anisotropies and differential streaming between different populations \citep{mars06,hellal06}.
While the solar wind electrons are relatively strongly collisional, 
ions are essentially collisionless, except in slow solar wind streams
\citep{neug76,mago83,lima86}. However, many ion properties
in the solar wind are relatively well ordered by the 
ion collisionality measured through an estimated collisional age \citep{kaspal08}.
On the other hand, as many solar wind properties are correlated \cite[cf.,][]{hetr14}
the role of collisions in the solar wind is not clear.

Transport in weakly collisional plasmas may strongly
deviate from theoretical predictions obtained for collision-dominated plasmas
\citep{spha53,brag65}. Even for the relatively strongly collisional
electrons such predictions (derived as a perturbation of thermal equilibrium) generally fail \cite[cf.,][]{landal14}.
The collisional energy and momentum transport coefficients can be calculated 
by taking moments of the collisional operator
assuming a particular form of particle distribution functions \citep{koga61,lehn67,basc81,hema85}.
For drifting bi-Maxwellian velocity distribution functions these
transport coefficients can be derived in
 in a closed form involving generalized double hypergeometric functions \citep{hetr09}.

For modeling of the solar wind, simplified versions of the transport coefficients
which neglect the temperature anisotropy and/or the differential streaming 
are often used \citep{hernal87,echial11,marual13,cran14,tracal15}. Observed important ion temperature
anisotropies and differential streaming make such approaches questionable
\cite[cf.,][]{mattal12}.
In this paper we quantify the collisional energy and momentum transport coefficients
in the solar wind using WIND/SWE observations at 1 AU. 
This paper is organized as follows: section~\ref{transcoef} gives an overview
of theoretical collisional transport coefficients for drifting bi-Maxwellian velocity distribution functions;
in this section
the collisional isotropization, relative deceleration and heating frequencies are defined.
Section~\ref{data} presents an analysis of the WIND/SWE data for proton and alpha particles. The different
collision frequencies are evaluated and compared with each other. The error due to neglecting
 the temperature anisotropy and/or the differential streaming is estimated.
Section~\ref{discussion} summarizes and discusses the results.

\section{Transport coefficients}

\label{transcoef}

Here we assume a homogeneous plasma consisting of species with bi-Maxwellian velocity distribution
functions $f_{\mathrm{s}}$ drifting along the magnetic field (here subscripts s and t denote different species;
subscripts $\|$ and $\perp$ denote directions with respect to the ambient magnetic field):
\begin{equation}
f_{\mathrm{s}}(v_\|,v_\perp)= \frac{n_{\mathrm{s}}}{(2\pi)^{3/2} v_{\mathrm{s}\|}v_{\mathrm{s}\perp}^2} \mathrm{e}^{-\frac{v_\perp^2}{2 v_{\mathrm{s}\perp}^2}
     -\frac{(v_\|-v_{\mathrm{s}})^2}{2 v_{\mathrm{s}\|}^2}}
\end{equation}
where $n_{\mathrm{s}}$ is the species number density,
 $v_{\mathrm{s}\|}=(k_B T_{\mathrm{s}\|}/m_{\mathrm{s}})^{1/2}$ and
$v_{\mathrm{s}\perp}=(k_B T_{\mathrm{s}\perp}/m_{\mathrm{s}})^{1/2}$
 are the parallel and perpendicular thermal velocities
corresponding to the parallel and perpendicular temperatures, $T_{\mathrm{s}\|}$ and $T_{\mathrm{s}\perp}$, 
respectively; $k_B$
is the Boltzmann constant,  $m_{\mathrm{s}}$ is the mass,
and $v_{\mathrm{s}}$ is the parallel drift velocity.
For these distribution functions the collisional transport coefficients for the parallel and perpendicular temperatures
 may be given as \cite[cf.,][]{hetr09}
\begin{align}
\left(\frac{\mathrm{d} T_{\mathrm{s}\perp}}{\mathrm{d}t}\right)_{{c}}
&= -\nu_{T\mathrm{s}} \left( T_{\mathrm{s}\perp}-T_{\mathrm{s}\|} \right)
+ T_{\mathrm{s}\perp} \sum\limits_{\mathrm{t}\ne \mathrm{s}} \nu_{H\mathrm{s}\perp}^{(\mathrm{t})} \\
\left(\frac{\mathrm{d} T_{\mathrm{s}\|}}{\mathrm{d}t}\right)_{{c}}
&= 2\nu_{T\mathrm{s}} \left( T_{\mathrm{s}\perp}-T_{\mathrm{s}\|} \right)
+ T_{\mathrm{s}\|} \sum\limits_{\mathrm{t}\ne \mathrm{s}} \nu_{H\mathrm{s}\|}^{(\mathrm{t})}
\end{align}
where $\nu_{T\mathrm{s}}$ is the (intraspecies) isotropization frequency \cite[cf.][]{koga61}
\begin{equation}
\nu_{T\mathrm{s}}=
\frac{ q_{\mathrm{s}}^4 n_{\mathrm{s}} \ln\Lambda_{\mathrm{ss}}}
{30\pi^{3/2}\epsilon_0^2 m_{\mathrm{s}}^2 v_{\mathrm{s}\|}^3 }
\ _2F_1
\left(\begin{array}{c|}2,3/2\\ 7/2\end{array}\,1-A_{\mathrm{s}}\right)
\label{nut}
\end{equation}
and $_2F_1$ is the standard (Gauss) hypergeometric function.
Here $\epsilon_0$ denotes the electric permittivity,
$q_{\mathrm{s}}$  is the  charge;
$A_{\mathrm{s}}=T_{\mathrm{s}\perp}/T_{\mathrm{s}\|}$ is the temperature
anisotropy,
and $\ln\Lambda_{\mathrm{ss}}$ is the Coulomb
logarithm.

The energy transfer between the different species is quantified by the perpendicular and parallel heating rates
$\nu_{H\mathrm{s}\perp}^{(\mathrm{t})}$ and  $\nu_{H\mathrm{s}\|}^{(\mathrm{t})}$
which may be expressed analytically as
\begin{align}
\nu_{H\mathrm{s}\perp}^{(\mathrm{t})}&=\frac{\nu_{\mathrm{st}}}{A_{\mathrm{st}}}\left[\frac{m_{\mathrm{st}}}{m_{\mathrm{t}}}\left(\frac{T_{\mathrm{t}\perp}}{T_{\mathrm{s}\perp}}-1\right)F_{2\frac{1}{2}\frac{5}{2}}^{(\mathrm{st})}+F_{2\frac{1}{2}\frac{5}{2}}^{(\mathrm{st})}-F_{1\frac{1}{2}\frac{5}{2}}^{(\mathrm{st})}\right] \label{nuhper}\\
\nu_{H\mathrm{s}\|}^{(\mathrm{t})}	&=\nu_{\mathrm{st}}
\bigg[\frac{m_{\mathrm{st}}}{m_{\mathrm{t}}}\left(\frac{T_{\mathrm{t}\|}}{T_{\mathrm{s}\|}}-1\right)F_{1\frac{1}{2}\frac{5}{2}}^{(\mathrm{st})}
-2\left(F_{2\frac{1}{2}\frac{5}{2}}^{(\mathrm{st})}-F_{1\frac{1}{2}\frac{5}{2}}^{(\mathrm{st})}\right) \nonumber \\
&\ \ \ +\frac{v_{\mathrm{st}}^{2}}{2v_{\mathrm{st}\|}^{2}}F_{1\frac{3}{2}\frac{5}{2}}^{(\mathrm{st})} \label{nuhpar}
\bigg].
\end{align}  
Here
\begin{equation}
v_{\mathrm{st}\|}=\sqrt{\frac{v_{\mathrm{s}\|}^2+v_{\mathrm{t}\|}^2}{2}} \ \ \textrm{and} \ \ 
v_{\mathrm{st}\perp}=\sqrt{\frac{v_{\mathrm{s}\perp}^2+v_{\mathrm{t}\perp}^2}{2}}
\end{equation}
are combined parallel and
perpendicular thermal velocities, respectively,
$v_{\mathrm{st}} = v_{\mathrm{s}}-v_{\mathrm{t}}$ is the relative velocity between the two species,
$m_{\mathrm{st}}=m_{\mathrm{s}} m_{\mathrm{t}}/(m_{\mathrm{s}}+m_{\mathrm{t}})$
is a combined mass
\begin{equation}
A_{\mathrm{st}}=\frac{v_{\mathrm{st}\perp}^2}{v_{\mathrm{st}\|}^2}= \frac{m_{\mathrm{t}} T_{\mathrm{s}\perp}+m_{\mathrm{s}} T_{\mathrm{t}\perp}}
       { m_{\mathrm{t}} T_{\mathrm{s}\|}+m_{\mathrm{s}} T_{\mathrm{t}\|}}
\end{equation}
 is a combined temperature anisotropies,
and
\begin{equation}
\nu_{\mathrm{st}}= \frac{ q_{\mathrm{s}}^2 q_{\mathrm{t}}^2 n_{\mathrm{t}}}  
{12\pi^{3/2}\epsilon_0^2 m_{\mathrm{s}} m_{\mathrm{st}} v_{\mathrm{st}\|}^3 } \ln\Lambda_{\mathrm{st}}
\end{equation}
is a collision frequency of species s on species t ($ \ln\Lambda_{\mathrm{st}}$ being the corresponding Coulomb logarithm).
Finally, $F_{abc}^{(\mathrm{st})}$ are defined through
 generalized double hypergeometric or
 Kamp\'e de F\'eriet functions \citep{exto76}
\begin{align}
F_{abc}^{(\mathrm{st})}=\mathrm{e}^{-\frac{v_{\mathrm{st}}^{2}}{4v_{\mathrm{st}\|}^{2}}}F_{1\cdot1}^{2\cdot\cdot}\left(\begin{array}{c|}
a,b\\
c;b
\end{array}\, 1-A_{\mathrm{st}},A_{\mathrm{st}}\frac{v_{\mathrm{st}}^{2}}{4v_{\mathrm{st}\|}^{2}}\right).
\end{align}
These functions can be represented as double series
\begin{equation}
F^{2\cdot\cdot}_{1\cdot1}
\left(\begin{array}{c|} a,b \\ c;d \end{array}\,
 x,y\right)=
\sum\limits_{n,k=0}^{\infty}
  \frac{ (a)_{n+k} (b)_{n+k} }{ (c)_{n+k} (d)_k }
    \frac{x^n }{n!}  \frac{y^k} {k!}.
\label{f211}
\end{equation}
where $(a)_n$ denotes the Pochhammer symbol
\begin{equation}
 (a)_n= \frac{\Gamma(a + n)}{\Gamma(a)} = a(a + 1) \ldots (a + n - 1),
\end{equation}
$\Gamma$ being the gamma function.
These series are absolutely convergent
for any $y$ and for $|x|<1$. Outside this region an analytic continuation is
needed. For the special case $b=d$ needed here, there exists a simple integral representation \citep{hetr09}:
\begin{align}
F^{2\cdot\cdot}_{1\cdot1}
\left(\begin{array}{c|} a,b \\ c;b \end{array}\,
 x,y\right) 
&= \frac{\Gamma(c)}{\Gamma(a)\Gamma(c-a)}\label{intrep} \\ 
&\times \int\limits_{0}^{1} \frac{ t^{a-1} (1-t)^{c-a-1}}{(1-t x)^b} \mathrm{e}^{\frac{ty}{1-tx}} \mathrm{d}t.
\nonumber 
\end{align}
which may be used for numerical evaluation.

We also define the mean heating rate $\nu_{H\mathrm{s}}^{(\mathrm{t})}$
as
\begin{equation}
\nu_{H\mathrm{s}}^{(\mathrm{t})}=\frac{1}{3}\frac{T_{\mathrm{s}\|}}{T_{\mathrm{s}}}
\nu_{H\mathrm{s}\|}^{(\mathrm{t})}
+ \frac{2}{3} \frac{T_{\mathrm{s}\perp}}{T_{\mathrm{s}}}
\nu_{H\mathrm{s}\perp}^{(\mathrm{t})}
\end{equation}
where $T_{\mathrm{s}}=(2 T_{\mathrm{s}\perp}+T_{\mathrm{s}\|} )/3$ is the mean temperature of the species.

For the relative deceleration between 
species s and t through Coulomb collisions,
one gets \cite{hetr09}
\begin{align}
\left(\frac{\mathrm{d}v_{\mathrm{st}}}{\mathrm{d}t}\right)_{{c}}^{(\mathrm{st})}	=-\nu_{V}^{(\mathrm{st})} v_{\mathrm{st}} 
\end{align}
where the deceleration frequency $\nu_{V}^{(\mathrm{st})}$ may be given as
\begin{align}
\nu_{V}^{(\mathrm{st})}	=\frac{q_{\mathrm{s}}^{2}q_{\mathrm{t}}^{2}n_{\mathrm{st}}}{24\pi^{3/2}\epsilon_{0}^{2}m_{\mathrm{st}}^{2}v_{\mathrm{st}\|}^{3}}\ln\Lambda_{\mathrm{st}}F_{1\frac{3}{2}\frac{5}{2}}^{(\mathrm{st})}
\end{align} 
where
$n_{\mathrm{st}}=  (n_{\mathrm{s}}m_{\mathrm{s}}+n_{\mathrm{t}}m_{\mathrm{t}})/(m_{\mathrm{s}}+m_{\mathrm{t}})$
is a combined number density.

\section{Collisional transport coefficients}
\label{data}
Here we use fitted data
from the two Faraday Cup instruments in the Solar Wind Experiment (SWE) on the Wind spacecraft.
 WIND is a rotating spacecraft with a spin-axis
perpendicular to the ecliptic plane and a period of three seconds.
   A Faraday Cup is an energy/charge instrument with a
large, conical field of view  which
measures the current produced by particles within a given energy
window.
Proton and alpha particle properties, number densities and 
parallel and perpendicular temperatures are obtained
using a non-linear least-squares fitting of data to a theoretical model assuming 
 bi-Maxwellian proton and alpha particle distribution functions and the magnetic field 
direction obtained from three-second measurements provided by the Magnetic Field Investigation
 on the Wind spacecraft. In this paper
we use a large statistical data set (about 4 millions data points)
 from 1995 to 2012 \citep{kaspal08,marual12}.
We use only the data when the Wind spacecraft was situated in the solar wind at
about 1 AU (portions of time when it was
inside the magnetosphere before 2004 were removed from the data set).

For the fitted proton and alpha particle parameters we calculate the different transport coefficients 
(given in section~\ref{transcoef}) 
approximating
the ion-ion Coulomb logarithm by \protect\cite[cf.,][]{tracal15}
\begin{equation}
\ln\Lambda_{\mathrm{st}}=29.9 - \mathrm{ln}\left[\frac{q_\mathrm{s} q_\mathrm{t}\left(m_\mathrm{s}+m_\mathrm{t}\right) }
{e^3 \left(m_\mathrm{s}\tilde{T}_\mathrm{t}+m_\mathrm{t} \tilde{T}_\mathrm{s}\right)} \left(\frac{n_\mathrm{s} q_\mathrm{s}^2}{ \tilde{T}_\mathrm{s}}+
\frac{n_\mathrm{t} q_\mathrm{t}^2}{ \tilde{T}_\mathrm{t}}\right)^{1/2} \right]
\label{lambdast}
\end{equation}
where the temperature anisotropies and differential streaming are neglected.
In this expression $\tilde{T}_\mathrm{s}$ stands for the temperature of species s in electrovolts
and $e$ is the proton charge. For numerical
evaluation of the generalized double hypergeometric functions we use the integral representation,
Eq.~(\ref{intrep}).

We start with the proton-alpha particle deceleration frequency $\nu_V^{(\alpha\mathrm{p})}$ (henceforth
we drop the superscript)
with respect to the characteristic transit/expansion time $t_e=R/v_{sw}$ where
$R$ is radial distance from the sun (being 1 AU here) and $v_{sw}$ is the solar wind velocity.
The product of a collisional transport coefficient such as $\nu_V$ and the
expansion time $t_e$ may be used as a proxy for the collisional age \cite[cf.,][]{kaspal08},
defined as an integral value over a relevant time interval ($t_0,t$) of a given collisional frequency $\nu$ 
$A_c=\int_{t_0}^{t} \nu(t^\prime) \mathrm{d}t^\prime$
\cite[cf.,][]{saleal03,chhial16};
here we prefer to interpret $\nu_V t_e$ (and other such products) as a way to compare the two local characteristic
times.

\begin{figure}[htb]
\centerline{\includegraphics[width=8cm]{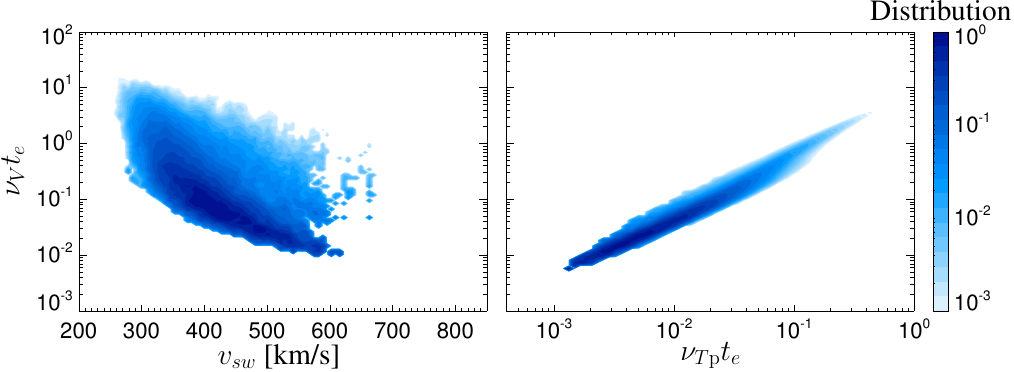}}
\caption{Proton-alpha particle deceleration frequency: Color scale plots of the relative frequency of ($v_{sw}$, $\nu_V t_e $)
(left panel) and of ($\nu_{T\mathrm{p}} t_e$, $ \nu_{V} t_e $) (right panel).
\label{nutv}}
\end{figure}

The left panel of Figure~\ref{nutv} shows the
distribution of data in the space ($v_{sw}$, $\nu_V t_e $).
This distribution (and all the following ones) was obtained by calculating number
of data points in each bin and dividing it by the bin size \cite[cf.,][]{marual12}; the results
are then globally renormalized to have the maximum value of the distribution equal to 1.
We recover the well known result that slower streams are typically more collisional.
It is interesting to compare the relative deceleration frequency $\nu_V$ and
the (proton-proton) isotropization frequency $\nu_{T\mathrm{p}}$
The right panel of Figure~\ref{nutv} shows the
distribution of data in the space ($\nu_{T\mathrm{p}} t_e$, $ \nu_{V} t_e $)
\cite[cf.,][]{hetr14}. The two collisional frequencies are almost proportional 
to each other as expected. Note that the deceleration frequency $\nu_V$ 
is noticeably larger than the proton isotropization one $\nu_{T\mathrm{p}}$ \cite[cf.,][]{mattal12}.

\begin{figure}[htb]
\centerline{\includegraphics[width=8cm]{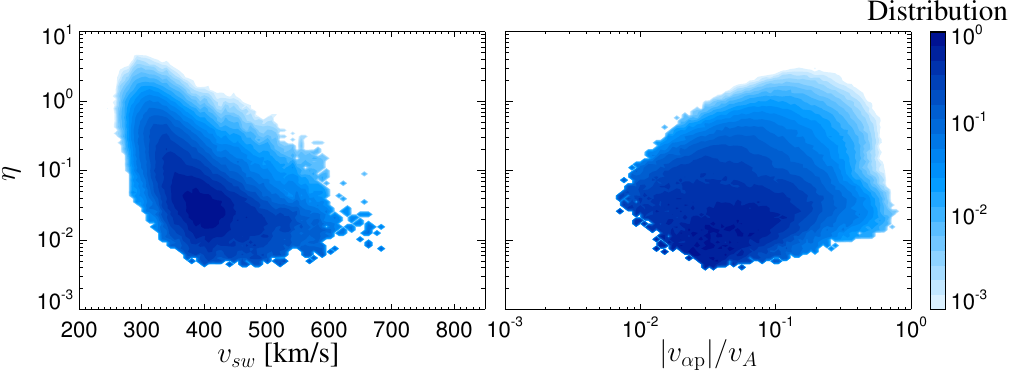}}
\caption{Proton-alpha deceleration with respect to the Alfv\'en velocity $v_A$: 
Color scale plots of the relative frequency of ($v_{sw}$, $ \eta$)
(left panel) and of ($ |v_{\mathrm{\alpha p}}| / v_A$, $ \eta$)
 (right panel).
\label{nuvva}}
\end{figure}

The relative proton-alpha particle velocity in the solar wind tends 
to decrease with the radial distance at a rate similar to that of the Alfv\'en velocity $v_A$
\citep{marsal82,versal15}.
It it therefore interesting to compare the collisional rate $({\mathrm{d}v_{\mathrm{\alpha p}}}/{\mathrm{d}t})_{{c}}^{(\mathrm{\alpha p})}$ and the decrease rate ${\mathrm{d}v_{A}}/{\mathrm{d}t}= v_{sw} {\mathrm{d}v_{A}}/{\mathrm{d}R} $;
let us investigate the ratio of the two rates:
\begin{equation}
\eta = \left| \frac{ \nu_{V} v_{\mathrm{\alpha p}}  }{ v_{sw} {\mathrm{d}v_{A}}/{\mathrm{d}R} } \right|.
\end{equation}
For the evaluation of $\eta$ we assume that the ion number densities decreases as $R^{-2}$ and
that the magnetic field follows the Parker spiral with the angle $45^{\mathrm{o}}$ at 1 AU.
The obtained results are shown in Figure~\ref{nuvva} where the left panel
shows the
distribution of data in the space ($v_{sw}$, $\eta$) whereas
the right panels the
distribution in ( $ |v_{\mathrm{\alpha p}}| / v_A$ , $\eta$).
 Figure~\ref{nuvva} indicate that Coulomb collisions between protons and alpha parties
 may be sufficient to decelerate the two species with respect to each other at 
a rate comparable to the decrease rate of $v_{A}$ in some slow solar wind streams. 

For the alpha particle isotropization frequency $\nu_{T\alpha}$ one expects much smaller
values than for the proton one owing to the smaller alpha particle abundance (and
typically larger temperatures). The observations, indeed, show that $\nu_{T\alpha}$ is
about an order of magnitude smaller than (and almost proportional to) $\nu_{T\mathrm{p}}$.

\begin{figure}[htb]
\centerline{\includegraphics[width=8cm]{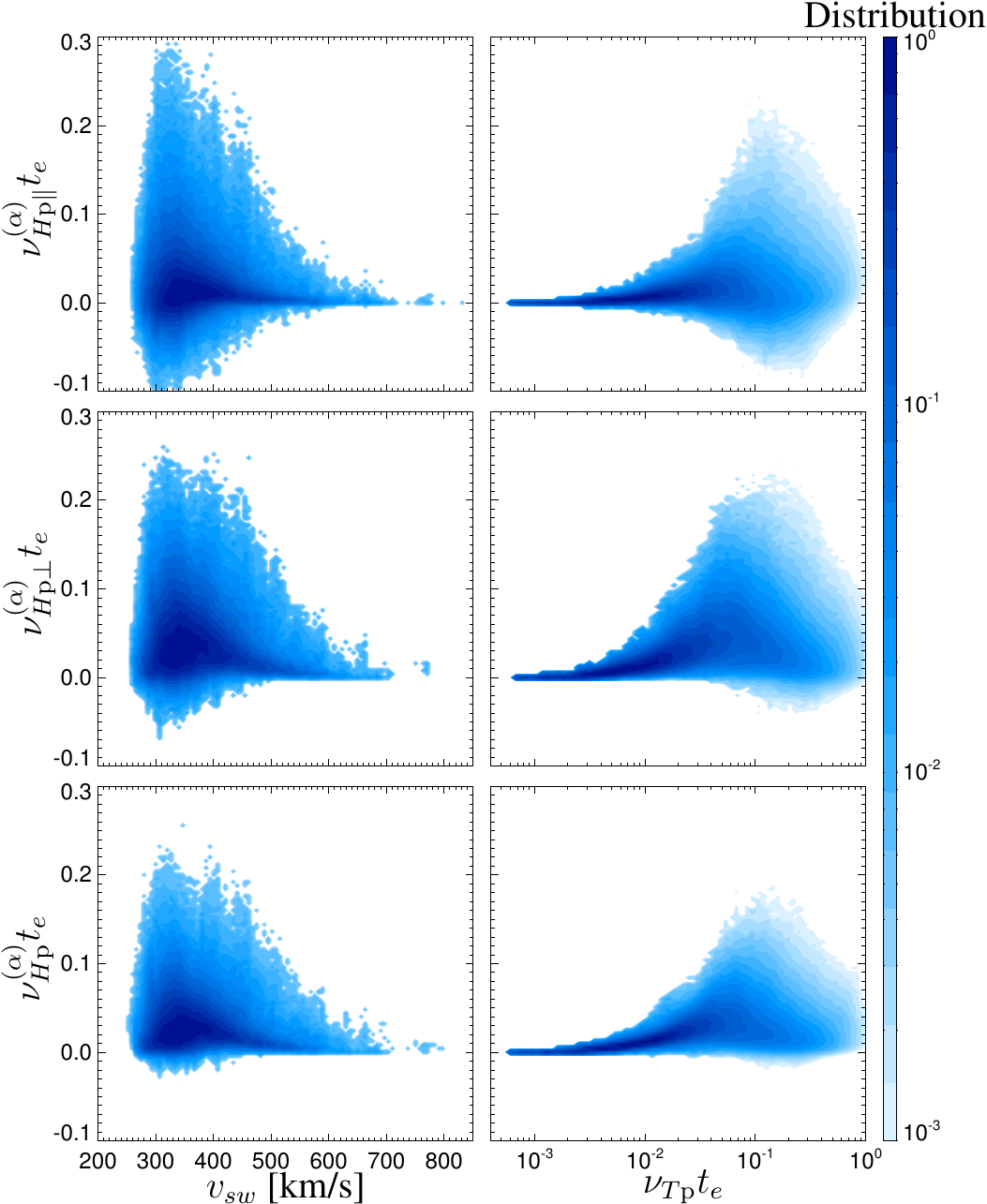}}
\caption{Proton heating rates through collisions with alpha particles: Color scale plots of the data distributions
in ($v_{sw}$, $\nu_{H\mathrm{p}\|}^{(\alpha)} t_e $)
(top left panel), ($\nu_{T\mathrm{p}} t_e$, $ \nu_{H\mathrm{p}\|}^{(\alpha)} t_e $) (top right panel),
($v_{sw}$, $\nu_{H\mathrm{p}\perp}^{(\alpha)} t_e $)
(middle left panel), ($\nu_{T\mathrm{p}} t_e$, $ \nu_{H\mathrm{p}\perp}^{(\alpha)} t_e $) (middle right panel),
($v_{sw}$, $\nu_{H\mathrm{p}}^{(\alpha)} t_e $)
(bottom left panel), and in ($\nu_{T\mathrm{p}} t_e$, $ \nu_{H\mathrm{p}}^{(\alpha)} t_e $) (bottom right panel).
\label{nuh}}
\end{figure}

Let's now look at the heating rates due to the Coulomb collisions between protons and alpha particles.
Figure~\ref{nuh} shows the data distribution 
in ($v_{sw}$, $\nu_{H\mathrm{p}\|}^{(\alpha)} t_e $)
(top left panel), ($\nu_{T\mathrm{p}} t_e$, $ \nu_{H\mathrm{p}\|}^{(\alpha)} t_e $) (top right panel),
($v_{sw}$, $\nu_{H\mathrm{p}\perp}^{(\alpha)} t_e $)
(middle left panel),  ($\nu_{T\mathrm{p}} t_e$, $ \nu_{H\mathrm{p}\perp}^{(\alpha)} t_e $) (middle right panel),
($v_{sw}$, $\nu_{H\mathrm{p}}^{(\alpha)} t_e $)
(bottom left panel), and in ($\nu_{T\mathrm{p}} t_e$, $ \nu_{H\mathrm{p}}^{(\alpha)} t_e $) (bottom right panel).
Figure~\ref{nuh} indicates that protons (in slower, more collisional streams) are typically heated (in total) through collisions with alpha particles.
This is quite natural as alpha particles are usually hotter than protons and collisions tend to remove this
difference. Protons are sometimes cooled in either parallel or perpendicular directions; this happens
typically in the cases of parallel or perpendicular proton temperature anisotropy. Proton collisions
with alpha particles also tend to reduce the proton temperature anisotropy.

\begin{figure}[htb]
\centerline{\includegraphics[width=8cm]{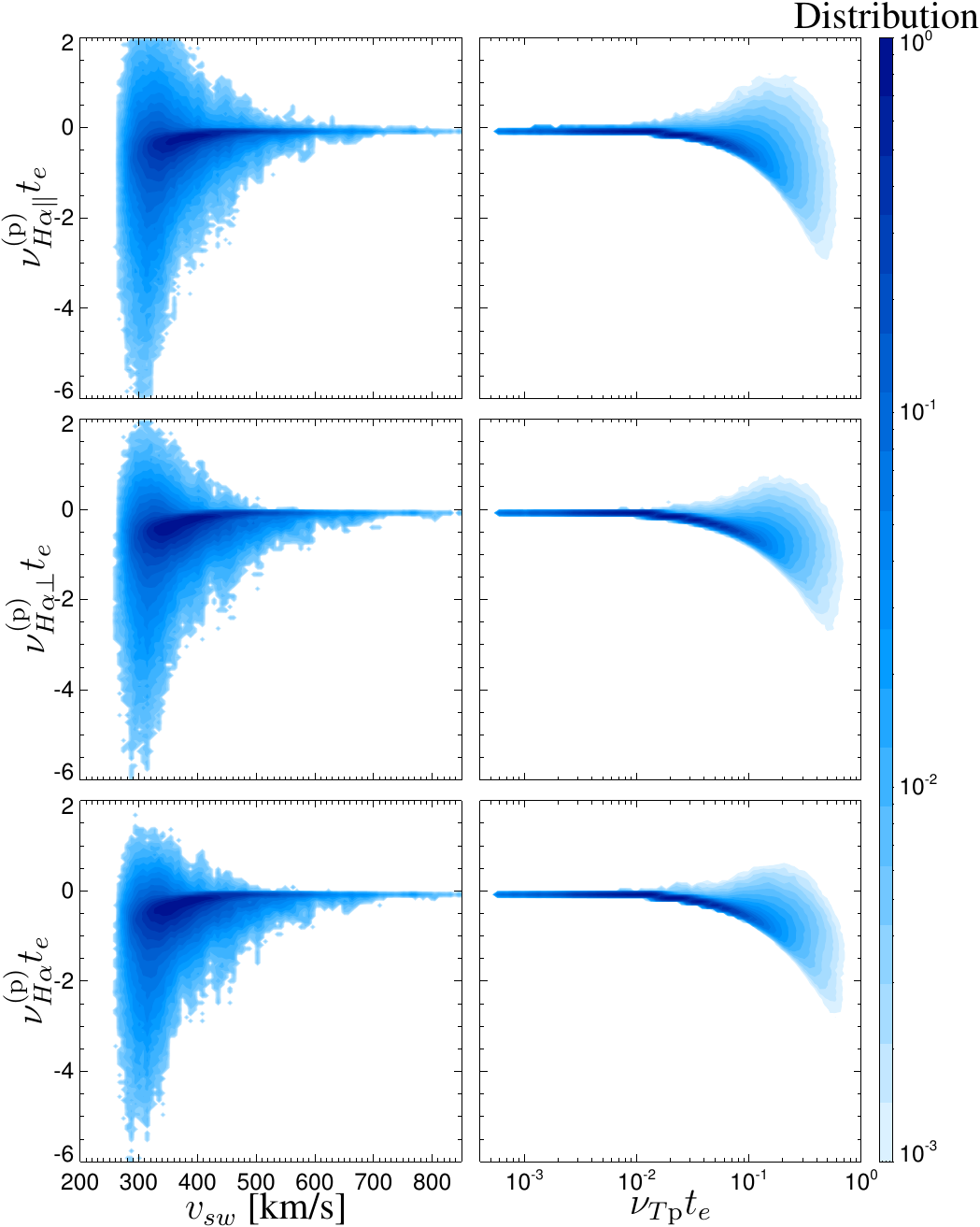}}
\caption{Alpha particle heating rates through collisions with protons: Color scale plots of the data distributions
($v_{sw}$, $\nu_{H\alpha\|}^{(\mathrm{p})} t_e $)
(top left panel), ($\nu_{T\mathrm{p}} t_e$, $ \nu_{H\alpha\|}^{(\mathrm{p})} t_e $) (top right panel),
($v_{sw}$, $\nu_{H\alpha\perp}^{(\mathrm{p})} t_e $)
(middle left panel), ($\nu_{T\mathrm{p}} t_e$, $ \nu_{H\alpha\perp}^{(\mathrm{p})} t_e $) (middle right panel),
($v_{sw}$, $\nu_{H\alpha}^{(\mathrm{p})} t_e $)
(bottom left panel), and in ($\nu_{T\mathrm{p}} t_e$, $ \nu_{H\alpha}^{(\mathrm{p})} t_e $) (bottom right panel).
\label{nuha}}
\end{figure}

Figure~\ref{nuha} shows the data distribution  in
($v_{sw}$, $\nu_{H\alpha\|}^{(\mathrm{p})} t_e $)
(top left panel), ($\nu_{T\mathrm{p}} t_e$, $ \nu_{H\alpha\|}^{(\mathrm{p})} t_e $) (top right panel),
($v_{sw}$, $\nu_{H\alpha\perp}^{(\mathrm{p})} t_e $)
(middle left panel), ($\nu_{T\mathrm{p}} t_e$, $ \nu_{H\alpha\perp}^{(\mathrm{p})} t_e $) (middle right panel),
($v_{sw}$, $\nu_{H\alpha}^{(\mathrm{p})} t_e $)
(bottom left panel), and in ($\nu_{T\mathrm{p}} t_e$, $ \nu_{H\alpha}^{(\mathrm{p})} t_e $) (bottom right panel).
Figure~\ref{nuha} indicates that alpha particles (in slower, more collisional streams)
are often cooled through collisions with protons; the energy stored in the proton-alpha particle
differential velocity (as well as a part of the alpha particle thermal energy) goes most probably to protons. 
The cases when alpha particles are heated correspond typically to the cases when protons and alpha particles
have comparable temperatures. In these cases the differential streaming energies is split to protons
and alpha particles. Alpha particles are sometimes heated in either parallel or perpendicular directions; this happens
typically in the cases of perpendicular or parallel alpha particle temperature anisotropy. Alpha particle collisions
with protons also tend to reduce the alpha particle temperature anisotropy.

The expressions for the different collisional transport coefficients are rather complex.
It is therefore interesting to test whether it is necessary to take into account
the temperature anisotropy or the differential streaming.
We estimated the importance of these parameters assuming isotropic populations
(with the mean temperature $T_\mathrm{s}=(T_{\mathrm{s}\|}+2T_{\mathrm{s}\perp})/3$)
and/or neglecting the differential ion streaming (i.e., setting $v_{\mathrm{p\alpha}}=0$)
for the present data set.
The relative error in the isotropization frequency $\nu_T$ is small ($\lesssim 10$ \%) when ignoring the
temperature anisotropy for protons and alpha particles.
However, the relative error in the relative deceleration frequency $\nu_V$ can be quite large ($\lesssim 40$ \%) 
when ignoring the
temperature anisotropy and/or the differential streaming. 
In the case of the (total) collisional heating rates $\nu_H$
the difference between the full (drifting \& anisotropic) version and that 
which neglects the temperature anisotropy and/or the differential streaming, $|\Delta\nu_H|$
could be of the order of $0.1/t_e$ for protons and $1/t_e$ for alpha particles (the parallel and perpendicular heating rates
have typically somewhat larger errors). These values are comparable to the maximum (absolute) values
of the heating rates (see Figures~\ref{nuh} and \ref{nuha}).

\section{Discussion}
\label{discussion}
We quantified proton and alpha particles collisional transport coefficients
at 1 AU using the WIND/SWE data. In agreement with previous studies our results
show that ion-ion Coulomb collisions 
are generally important for slow solar wind streams; they
tend to reduce the differences between the ion mean velocities and temperatures,
as well as the ion temperature anisotropies. The different collisional frequencies
have typically disparate values leading to different collisional ages.
The two isotropization frequencies and the relative deceleration frequency
are nearly proportional each to other so that one expects that the corresponding
collisional times will be also proportional each to other. On the other
hand, the heating rates are not simply related to these frequencies, and, moreover,
these heating rates have positive and negative values.

The observations indicate that the relative proton-alpha particle
collisional deceleration is more efficient than the proton isotropization
(and the alpha-particle isotropization) but, on the other hand, 
proton-alpha particle collisions also tend to reduce both the proton 
and alpha particle temperature anisotropies (in a relatively small
number of cases the proton-alpha particle collisions enhance the ion temperature anisotropy).
In some slow solar wind streams the Coulomb collisions are sufficiently strong
to reduce the differential velocity between proton and alpha particles
at a pace that is comparable to that of the decrease rate of the Alfv\'en velocity.

Protons are typically heated through collisions with alpha particles
whereas alpha particles are very often cooled; this is
a consequence of the temperature difference between the two species,
alpha particles are typically hotter than protons. When protons
and alpha particles have comparable temperatures both the species are
typically heated at the expense of the relative proton-alpha particle
velocity.
Our results indicate that the proton heating through collisions with alpha particles in slow solar wind streams 
reaches values $
\left({\mathrm{d} T_{\mathrm{p}}}/{\mathrm{d}t}\right)_{{c}}^{(\alpha)} 
 \sim 0.2 T_{\mathrm{p}}/t_e$ at 1 AU. This value is an important fraction of
the needed average proton heating rate estimated from the Helios observations for
an average slow solar wind ${\mathrm{d} T_{\mathrm{p}}}/{\mathrm{d}t} - (T_{\mathrm{p}}/{\mathrm{d}t})_\mathrm{CGL} 
 \sim 0.45 T_{\mathrm{p}}/t_e $
 \cite[cf.,][]{hellal13}. 
In the fast solar wind the needed proton heating rate has similar
values $\sim 0.32 T_{\mathrm{p}}/t_e $ but the collisional
proton heating rates are much smaller. Coulomb collisions with
alpha particles are not energetically important for protons in
the fast solar wind at 1 AU.

The collisional interaction of alpha particles with protons may
be important in slow solar wind streams. Similarly, \cite{tracal15} show
that Coulomb collision with protons may be also important for
other minor ions. It is possible that the (summed effect of the)
interaction between protons and other minor ions may have
a nonnegligible heating effect on protons; the abundances
of the other minor ions are much lower than the abundance
of alpha particles but together they have about 1 \% 
and they are about mass proportionally hotter than alpha particles.

Our results show that the different parallel and perpendicular
temperatures and differential velocities between ion species need to be taken into account for the collisional
transport coefficients especially in slow/collisional solar wind streams. 
For the given data set of the WIND/SWE observations at 1 AU neglecting
the ion temperature anisotropy and/or the differential streaming leads to relatively
large errors. Consequently, it is important to use the general, anisotropic
and drifting approximation for the collisional
transport coefficients, particularly for modeling of slow solar wind streams
where the induced error cumulates.
The present results have (at best) theoretical uncertainties of the order 
of $1/\ln \Lambda_\mathrm{st}$. Equation~(\ref{lambdast}) gives the
uncertainties 4--5 \% for the ion-ion interaction. Equation~(\ref{lambdast})
is simplified, the temperature anisotropy and the
differential streaming are neglected. It would be interesting to extend
Eq.~(\ref{lambdast}) to include these effect, but, on the other hand,
the Coulomb logarithm is only weakly/logarithmically dependent on the plasma
parameters, so that we do not expect significant changes. 
On the other hand,
our model assumes bi-Maxwellian particle velocity distribution functions
drifting with respect each other along the ambient magnetic field.
If the ion distribution functions in the solar wind
strongly depart from this model, these collisional transport coefficients are likely not
applicable and further work is needed. For instance,
the solar wind protons often consist of two populations \citep{mars06};
such a velocity distribution function
 can be to some extent modelled as a superposition
of two drifting bi-Maxwellian velocity distribution functions and the present
model can be directly applied. 
 
\acknowledgments
PH acknowledges grant 15-10057S of the Grant Agency of the Czech Republic
and the project RVO:67985815.
The research leading to these results has received funding from the
European Commission's 7th Framework Programme under
 the grant agreement \#284515 (project-shock.eu).
Wind data were obtained from the NSSDC website
http://nssdc.gsfc.nasa.gov.

\end{document}